\documentclass[cits]{PoS}
\usepackage{subfigure}

\title{Recent results from the ALICE experiment on open heavy flavours in hadronic collisions at the LHC}

\ShortTitle{Recent results from the ALICE experiment on open heavy flavours}

\author{\speaker{Sarah Porteboeuf-Houssais }\\%
        Laboratoire de Physique Corpusculaire (LPC), Clermont Universit\'e, Universit\'e Blaise Pascal, CNRS-IN2P3, Clermont-Ferrand, France\\
        E-mail: \email{sarah@clermont.in2p3.fr}}

\author{for the ALICE collaboration}

\abstract{ALICE is the LHC experiment devoted to the study of the Quark-Gluon Plasma (QGP). To probe this high energy density state of strongly interacting matter expected to be produced in heavy-ion collisions at high energies, measurements performed in various systems (pp, p--Pb and Pb--Pb) are compared to each other. Heavy quarks are  produced in initial  hard partonic scatterings on a short time scale and participate in the subsequent evolution of the medium. This makes them sensitive probes of the QGP. With ALICE, open heavy flavours are studied using D mesons  (D$^0$, D$^+$, D$^{*+}$) reconstructed via their hadronic decay channels in the mid-rapidity region ($|y|<0.5$) and with heavy-flavour decay leptons reconstructed in the electronic (muonic) channel in the central rapidity region  $|y|<0.9$ (forward rapidity region $2.5<y<4$). We present ALICE measurements of the nuclear modification factor $R_{\rm{AA}}$ and the elliptic flow $v_2$ for D mesons and heavy-flavour decay leptons  in  Pb--Pb collisions at $\sqrt{s_{\rm{NN}}}=2.76$ TeV. The corresponding measurements  in p--Pb collisions at $\sqrt{s_{\rm{NN}}}=5.02$ TeV are also discussed. First results on the azimuthal correlations between heavy-flavour particles and hadrons are presented as well as the charged-particle multiplicity dependence of heavy-flavour production in pp collisions at $\sqrt{s}=7$ TeV.
}

\FullConference{52 International Winter Meeting on Nuclear Physics\\
                 27 - 31 January 2014\\
                 Bormio, Italy}

\begin{document}

\section{Introduction}

A phase transition from nuclear matter to a deconfined state of quarks and gluons is predicted by lattice quantum chromodynamics (lQCD) \cite{karsh} to occur under extreme conditions of energy density. Currently, the physics of the QGP is studied using ultra relativistic heavy ion collisions at RHIC \cite{brahms,phenix,phobos,star} and, at much higher energies, at the CERN Large Hadron Collider (LHC) \cite{first_results}. 

Open heavy flavours (HF) are hadrons containing charm or beauty quarks produced in hadronic collisions. Due to their large masses, the initial c (b) quarks are produced in hard scatterings in the initial stage of the collision. Their production cross sections in proton-proton (pp) collisions can be calculated by means of perturbative QCD calculations (pQCD). It is, then, essential to study HF production in pp collisions as a test of pQCD and as a reference measurement for p--Pb and Pb--Pb collisions. Further insight into charm and beauty production in hadronic collisions can be obtained by measuring the angular correlations between heavy-flavour hadrons and the charged particles produced in the collision, as well as by studying heavy-flavour production as a function of the multiplicity of produced particles. In a nuclear environment, HF production cross sections are modified by cold nuclear matter effects. To assess the importance of such effects, in the interpretation of Pb--Pb results, reference measurements are done in p--Pb collisions.    

In nucleus-nucleus (A--A) collisions, considered as an incoherent sum of nucleon interactions, HF production is expected to scale with the number of nucleon-nucleon collisions $N_{\rm{coll}}$. The nuclear modification factor $R_{\rm{AA}}$ is sensitive to any modification of this scaling and is defined as:  
\begin{equation}
R_{\rm{AA}}=\frac{\mathrm{d}N_{\mathrm{AA}}/\mathrm{d}p_{\mathrm{T}}}{<N_{\mathrm{coll}}> \mathrm{d}N_{\mathrm{pp}}/\mathrm{d}p_{\mathrm{T}}}=\frac{\mathrm{d}N_{\mathrm{AA}}/\mathrm{d}p_{\mathrm{T}}}{<T_{\mathrm{AA}}> \mathrm{d}\sigma_{\mathrm{pp}}/\mathrm{d}p_{\mathrm{T}}} \nonumber
\end{equation}
where the $p_{\mathrm{T}}$ differential yield in A--A collisions is compared to the pp yield scaled by the mean number of binary nucleon-nucleon collisions ($N_{\mathrm{coll}}$). $N_{\mathrm{coll}}$ and the overlap function  $T_{\mathrm{AA}}$ can be calculated with the Glauber model \cite{Glauber}. Heavy quarks, being produced in initial hard scatterings, experience, before hadronizing, the full evolution of the formed medium in Pb--Pb collisions. As they are sensitive to hot medium effects, they are considered to be sensitive QGP probes. Moving through a coloured medium, such as the QGP, quarks and gluons lose energy via radiative and collisional interactions. Parton energy loss ($\Delta E$) is proportional to the Casimir coupling (3 for gluons and 4/3 for quarks). In addition, small-angle gluon radiation is expected to be reduced for heavy quarks (dead cone effect) \cite{Dead_cone, test_dead_cone}. The two combined effects lead to a parton energy loss ordering: $\Delta E_{\mathrm{g}}>\Delta E_{\mathrm{uds}}>\Delta E_{\mathrm{c}}>\Delta E_{\mathrm{b}}$ which may be reflected in the nuclear modification factor: $R_{\mathrm{AA}}^{\pi}<R_{\mathrm{AA}}^{\mathrm{D}}<R_{\mathrm{AA}}^{\mathrm{B}}$.

In non-central heavy-ion collisions, the initial spatial anisotropy is converted into a particle momentum anisotropy due to different pressure gradients in and out of the reaction plane ($\psi_{\mathrm{RP}}$). The anisotropy is quantified via a Fourier expansion in azimuthal angle with respect to the reaction plane. The second Fourier coefficient $v_2$ is called elliptic flow \cite{ollitrault}. If heavy quarks suffer significant interactions with the medium constituents, a non-zero $v_2$ is expected at low $p_{\mathrm{T}}$. 

Section \ref{exp_data_sample} is devoted to the experimental apparatus description and data samples. We present ALICE results in pp collisions in section \ref{pp}. Reference observables measured in p--Pb collisions are presented in section \ref{p--Pb}. The nuclear modification factor and elliptic flow results in Pb--Pb collisions are discussed in section \ref{Pb--Pb}.

\section{Experimental apparatus and data samples}
\label{exp_data_sample}

The ALICE apparatus is the LHC dedicated experiment for heavy-ion collisions \cite{ALICE}. It consists of a central barrel covering the pseudo-rapidity range $|\eta|<0.9$ embedded in a solenoid magnet providing a 0.5 T field, a forward muon spectrometer covering the pseudo-rapidity range $-4 < \eta< -2.5$ and a set of small acceptance detectors at forward and backward rapidities for event characterization and triggering. Open heavy-flavour measurements can be performed in various decay channels and kinematic regions. At central rapidities, HF can be measured via the identification of electrons from semi-leptonic decays or via the full reconstruction of charm mesons through their hadronic decays: $\mathrm{D}^0 \to \mathrm{K}^-  \pi^+$, $\mathrm{D}^+ \to \mathrm{K}^-  \pi^+ \pi^+$, $\mathrm{D}^{*+} \to \mathrm{D}^0  \pi^+$, $\mathrm{D}^+_s \to \mathrm{K}^-  \mathrm{K}^+ \pi^+$. At forward rapidities HF are measured in the semi-muonic decay channel.

In the central rapidity region, tracking is performed combining information from the Inner Tracking System (ITS), the Time Projection Chamber (TPC) and the Time Of Flight (TOF) detectors. The ITS is composed of six cylindrical layers of silicon detectors. The high spatial resolution of the reconstructed hits, the low material budget and the small distance from the beam pipe of the innermost layer allow the measurement of the track impact parameter in the transverse plane with a resolution better than 75 $\mu$m for transverse momenta $p_{\mathrm{T}} > 1$ GeV/$c$. Charged hadrons are identified using information from the TPC and the TOF detectors. The TPC is used to measure the specific energy loss ($\mathrm{d}E/\mathrm{d}x$) for the particle identification (PID) of charged particles. PID is supplemented by the TOF detector which is equipped with Multi-gap Resistive Plate Chambers (MRPCs). For high momentum electrons, information from the Transition Radiation Detector (TRD) and the ElectroMagnetic Calorimeter (EMCal) is used to suppress hadron contamination. 

In the forward rapidity region, muons are identified and characterized using the muon spectrometer, which consists of a front absorber followed by a 3 $\mathrm{T} \cdot \mathrm{m}$ dipole magnet, five tracking chamber stations and a trigger system. Each tracking station is made of two planes of cathode pad chambers. The trigger system is made of two stations with two planes of resistive plate chambers.

 The recent results presented in these proceedings were obtained by analyzing Run 1 data for various systems (pp, p--Pb and Pb--Pb) and centre-of-mass energies per nucleon-nucleon pair from 2.76 TeV to 7 TeV. Information on data samples and trigger conditions are summarized in table \ref{data sample}.

\begin{table}
\begin{center}
\begin{tabular}{|c|c|c|c|c|c|}
\hline
system & Energy & LHC Run & D mesons & HF electrons & HF muons \\ 
\hline
\hline
pp      & 2.76 TeV & 2011 & 1.1 nb$^{-1}$                                                                               & 0.5 (11.9) nb$^{-1}$                                                     & 19 nb-1  \\
          &                &          &   MB trigger                                                                                    &  MB (EMCAL) trigger                                                       & muon trigger \\
          &                &          & \textcolor{red}{cross sections} \cite{charm_2.76}                     & \textcolor{green}{cross sections}                                  & \textcolor{red}{cross sections} \cite{decay_muon_2.76} \\
\hline
pp      & 7 TeV      & 2010 & 5 nb$^{-1}$                                                                                  & 2.6 nb$^{-1}$                                                                    & 16.5 nb-1  \\
          &                &          &   MB trigger (\cite{Ds_pp_7})                                                        &    MB trigger (\cite{HFE_pp_7})                                            & muon trigger \\
          &                &          &  \textcolor{red}{cross sections} \cite{central_charm_pp_7}       &  2.2 nb$^{-1}$                                                                 & \textcolor{red}{cross sections} \cite{HFmu_pp_7} \\
          &               &           &\textcolor{green}{D-h correlations}                                              &   \textcolor{red}{cross sections} \cite{efromB_7}          & \\       
          &               &           & \textcolor{green}{D vs. multiplicity }                                           &                                                                                           & \\       
\hline
p--Pb   & 5.02 TeV & 2013 & 48.6 $\mu$b$^{-1}$     MB trigger                                                                   & 48.6 $\mu$b$^{-1}$         MB trigger                                                       & work in progress \\
          &               &           &\textcolor{green}{$R_{\mathrm{pPb}}$}                                             &\textcolor{green}{$R_{\mathrm{pPb}}$}                                     & \\
          &               &           &                                                                                                       &  \textcolor{green}{e-h correlation}                                       & \\           
\hline 
Pb--Pb & 2.76 TeV & 2010 & 1.12 $\mu$b$^{-1}$      0-80\%                                                                   & 2  $\mu$b$^{-1}$   0-80\%                                                                    & 2.7 $\mu$b$^{-1}$  MB trigger\\
           &               &           &  \textcolor{red}{$R_{\mathrm{AA}}$} \cite{D_RAA}                          &                                                                                             &  \textcolor{red}{$R_{\mathrm{AA}}$} \cite{decay_muon_2.76}  \\           
\hline
Pb--Pb & 2.76 TeV & 2011 & 23 $\mu$ b$^{-1}$   in 0-10\%                                                     & 22 (37)  $\mu$b$^{-1}$  in 0-10\%                                     & 11.3 $\mu$b$^{-1}$ in  in 0-10\% \\
           &                &          &   6.2 $\mu$b$^{-1}$   in 10-50\%                                                & 6 (34) $\mu$b$^{-1}$ in 20-40\%                                       & 3.5 $\mu$b$^{-1}$ in  in 10-40\% \\       
           &                &          &                                                                                                       &  MB (EMCAL) trigger                                                              &  \\
           &               &           &  \textcolor{red}{$v_{\mathrm{2}}$}\cite{D_Pb--Pb_v2},  \textcolor{green}{$R_{\mathrm{AA}}$ }                      &   \textcolor{green}{$v_{\mathrm{2}}$ }, \textcolor{green}{$R_{\mathrm{AA}}$ }                    &  \textcolor{green}{$v_{\mathrm{2}}$}  \\                
\hline

\end{tabular}
\end{center}
\caption{Data samples used for open heavy-flavour measurements during Run 1. Red: published results discussed in these proceedings. Green: preliminary measurements discussed in these proceedings. In parenthesis: references to published results not discussed in these proceedings.}  
\label{data sample}     
\end{table}

\begin{figure}
\begin{center}
\subfigure{
	\label{fig:D-h_correlation}
	\includegraphics[width=0.430\columnwidth]{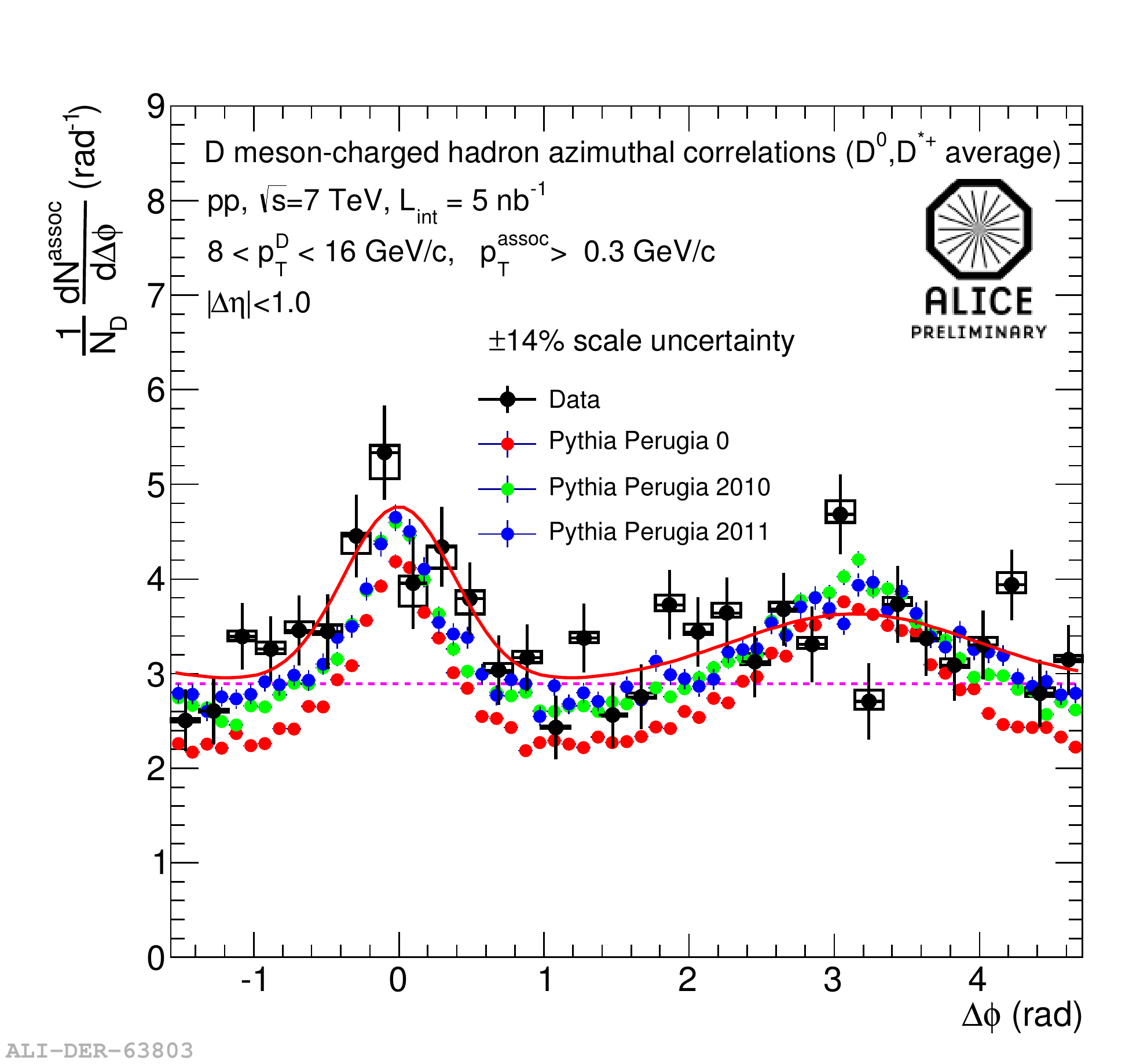}
	} 
\subfigure{
	\label{fig:D_vs_mult}
	\includegraphics[width=0.410\columnwidth]{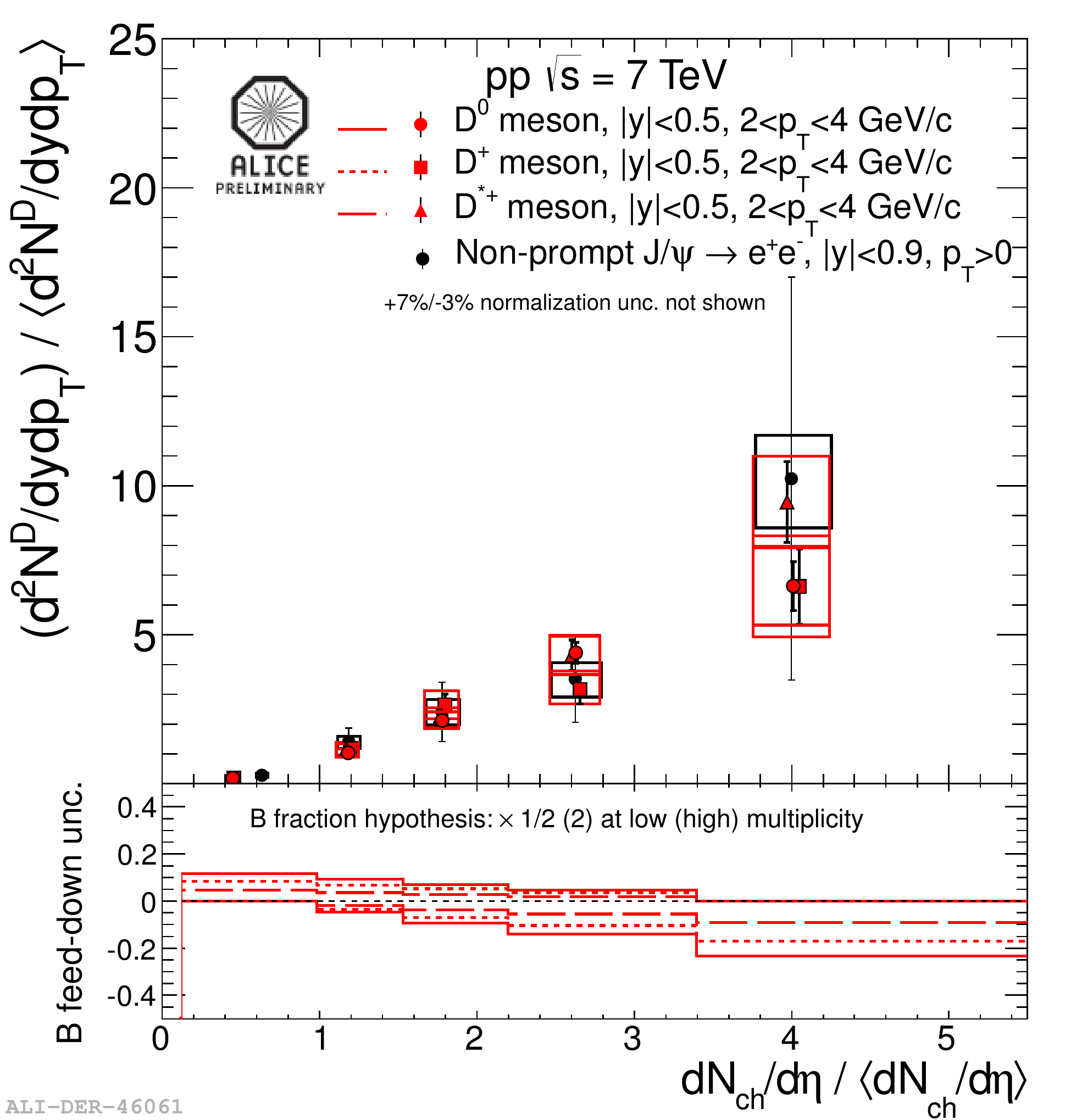}
	} 
\caption{Left: D meson-charged hadron azimuthal correlations in pp collisions at $\sqrt{s}=7$ TeV compared to PYTHIA 6.4.
	Right: D mesons and non prompt J/$\psi$ versus multiplicity in pp collisions at $\sqrt{s}=7$ TeV.
\label{fig:pp_exclusif}
}
\end{center}
\end{figure}

\section{Results from pp collisions at $\sqrt{s}= 2.76$ TeV and $7$ TeV}
\label{pp}

The $p_{\mathrm{T}}$-differential cross sections at $\sqrt{s} = 2.76$ TeV for open heavy flavours are not discussed in detail in these proceedings.  pQCD-based calculations \cite{FONLL, GM-VNFS} are compatible with data within uncertainties, in the measured $p_{\mathrm{T}}$ range ($p_{\mathrm{T}}$ < 12 GeV/$c$) for D mesons \cite{charm_2.76}, HF decay electrons \cite{decay_e_2.76} and HF decay muons \cite{decay_muon_2.76}. The same conclusion holds for $\sqrt{s} = 7$ TeV \cite{efromB_7,HFE_pp_7,HFmu_pp_7}, indicating a good understanding of HF production in pp collisions at LHC energies with pQCD.

More exclusive measurements, as, for example, the azimuthal correlations between D mesons and charged hadrons presented in Fig. \ref{fig:pp_exclusif}, have recently been achieved. These can give insight into the charm production mechanism in pp collisions via the associated hadron yields in the near and away side regions. Such a measurement is an important reference for studies of charm jet modification due to hot medium effects in Pb--Pb collisions. The measured correlation is described by PYTHIA 6.4 within statistical and systematic uncertainties. One should note that this measurement is performed multiplicity integrated, while non trivial effects in pp collisions (e.g the so-called ridge \cite{CMS_ridge}) only occur in high multiplicity events. More precise measurements are expected from Run 2. 

D-meson and non-prompt J/$\psi$ production as a function of the multiplicity of charged particles produced in the collision are shown in the right panel of Fig. \ref{fig:pp_exclusif}. This observable is related to the underlying pp event activity accompanying heavy-flavour production. The interpretation of the observed approximate linear increase is connected with multi-parton interactions and possible contributions from hadronic activity around HF production (fragmentation and final state radiation).

\begin{figure}
\begin{center}
\subfigure{
	\label{fig:D_RpPb}
	\includegraphics[width=0.410\columnwidth]{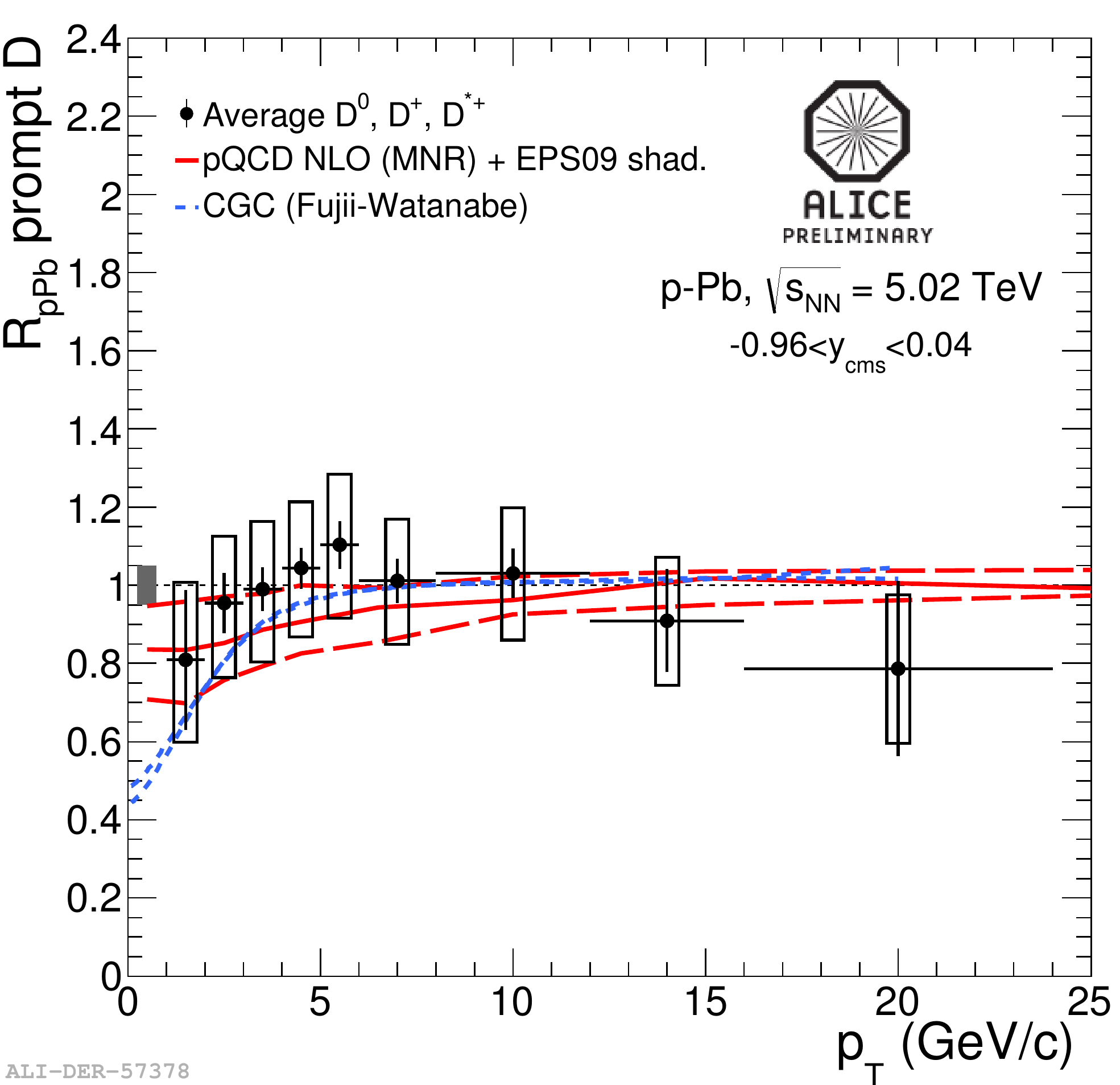}
	} 
\subfigure{
	\label{fig:HF_RpPb}
	\includegraphics[width=0.480\columnwidth]{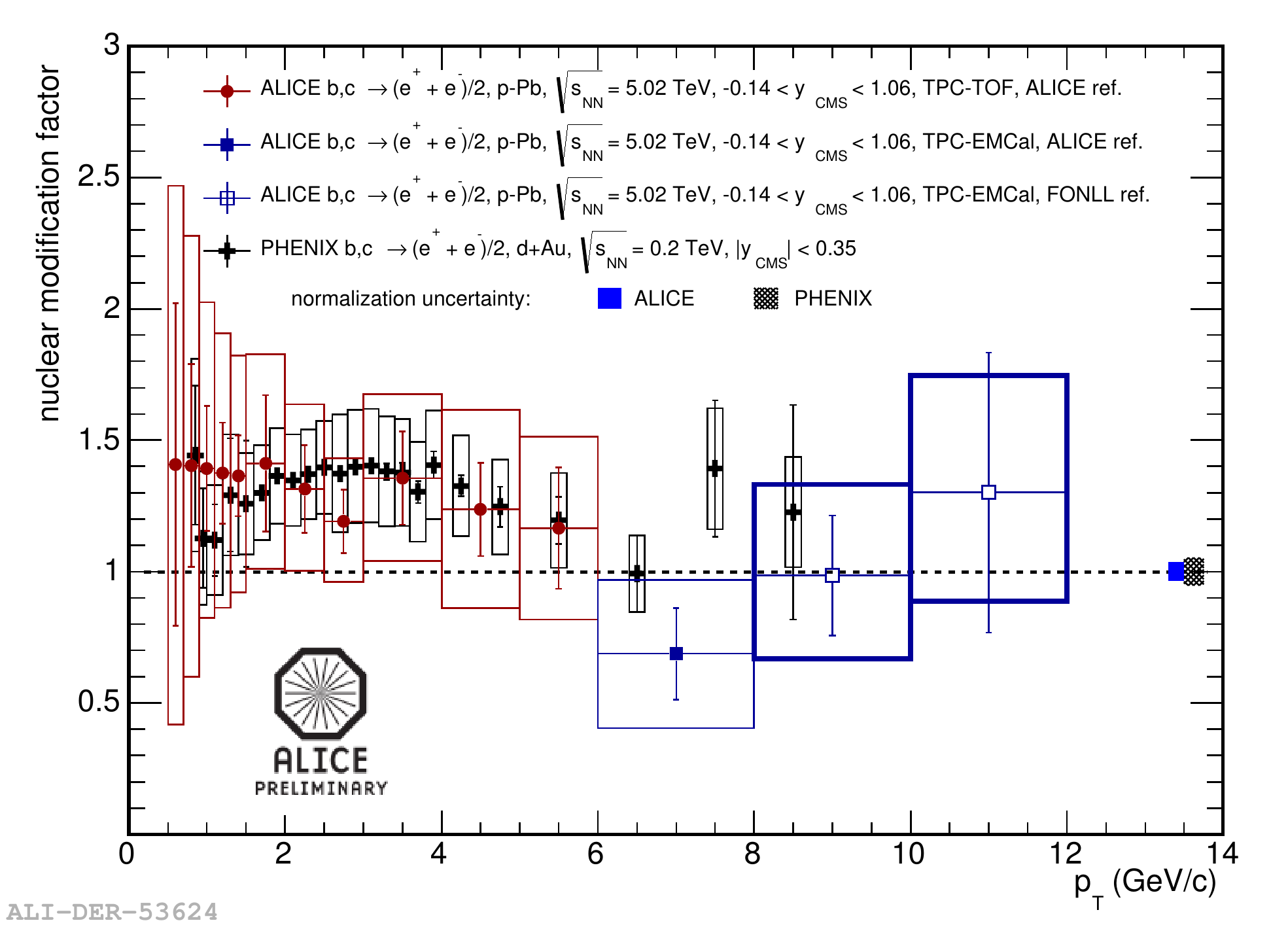}
	} 
\caption{D-meson and heavy-flavour decay electron $R_{\mathrm{pPb}}$ in p--Pb collisions at $\sqrt{s_{\mathrm{NN}}}=5.02$ TeV.
	Left: D-meson $R_{\mathrm{pPb}}$ as a function of $p_{\mathrm{T}}$ compared to theoretical predictions. 
	Right: HF decay electrons $R_{\mathrm{pPb}}$ as a function of $p_{\mathrm{T}}$ compared with the PHENIX measurement at $\sqrt{s_{\mathrm{NN}}}=200$ GeV.	
\label{fig:RpPb}
}
\end{center}
\end{figure}

\begin{figure}
\begin{center}
\subfigure{
	\label{fig:e-h_correlations_1D}
	\includegraphics[width=0.450\columnwidth]{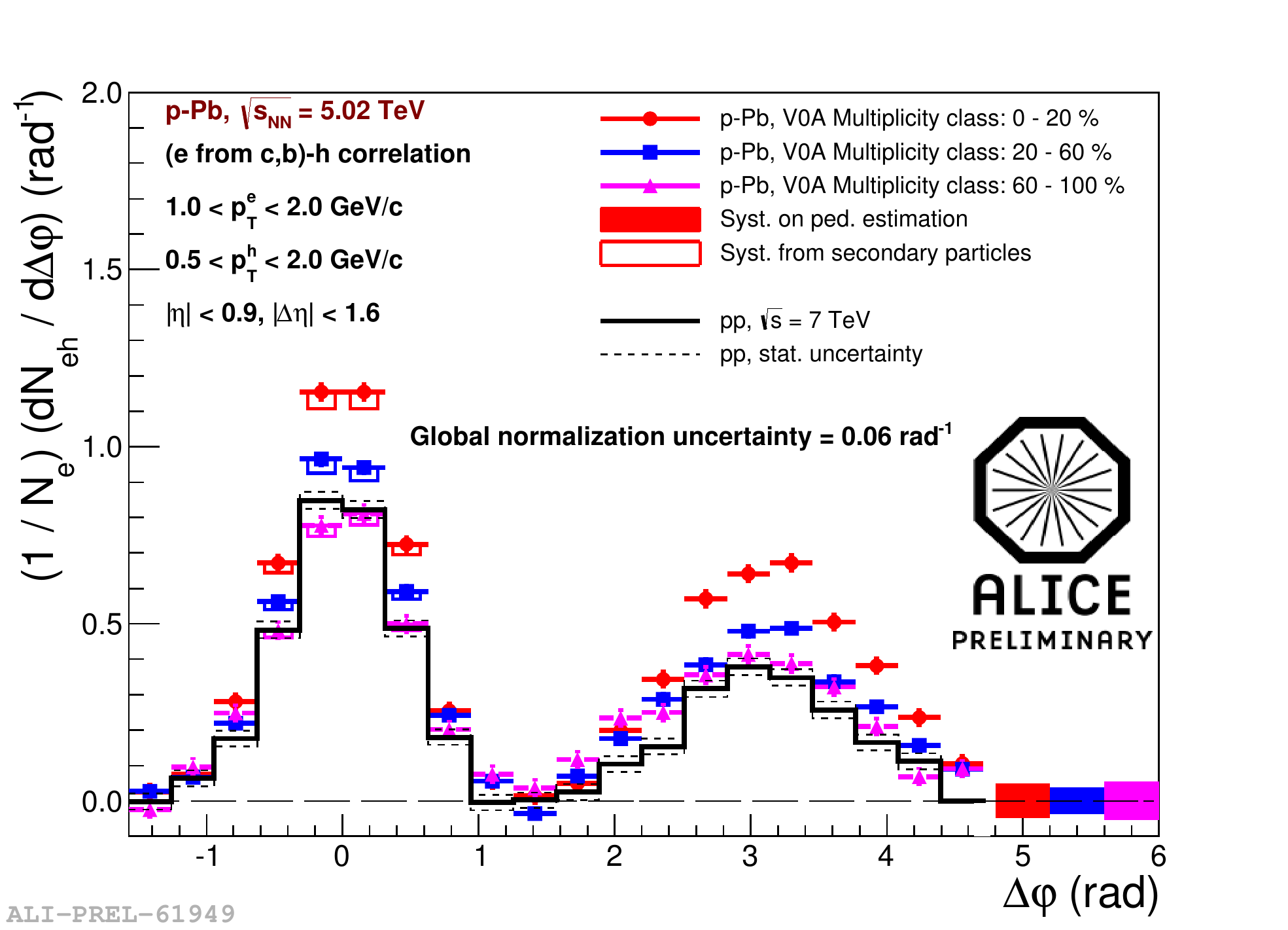}
	} 
\subfigure{
	\label{fig:e-h_correlations_2D}
	\includegraphics[width=0.450\columnwidth]{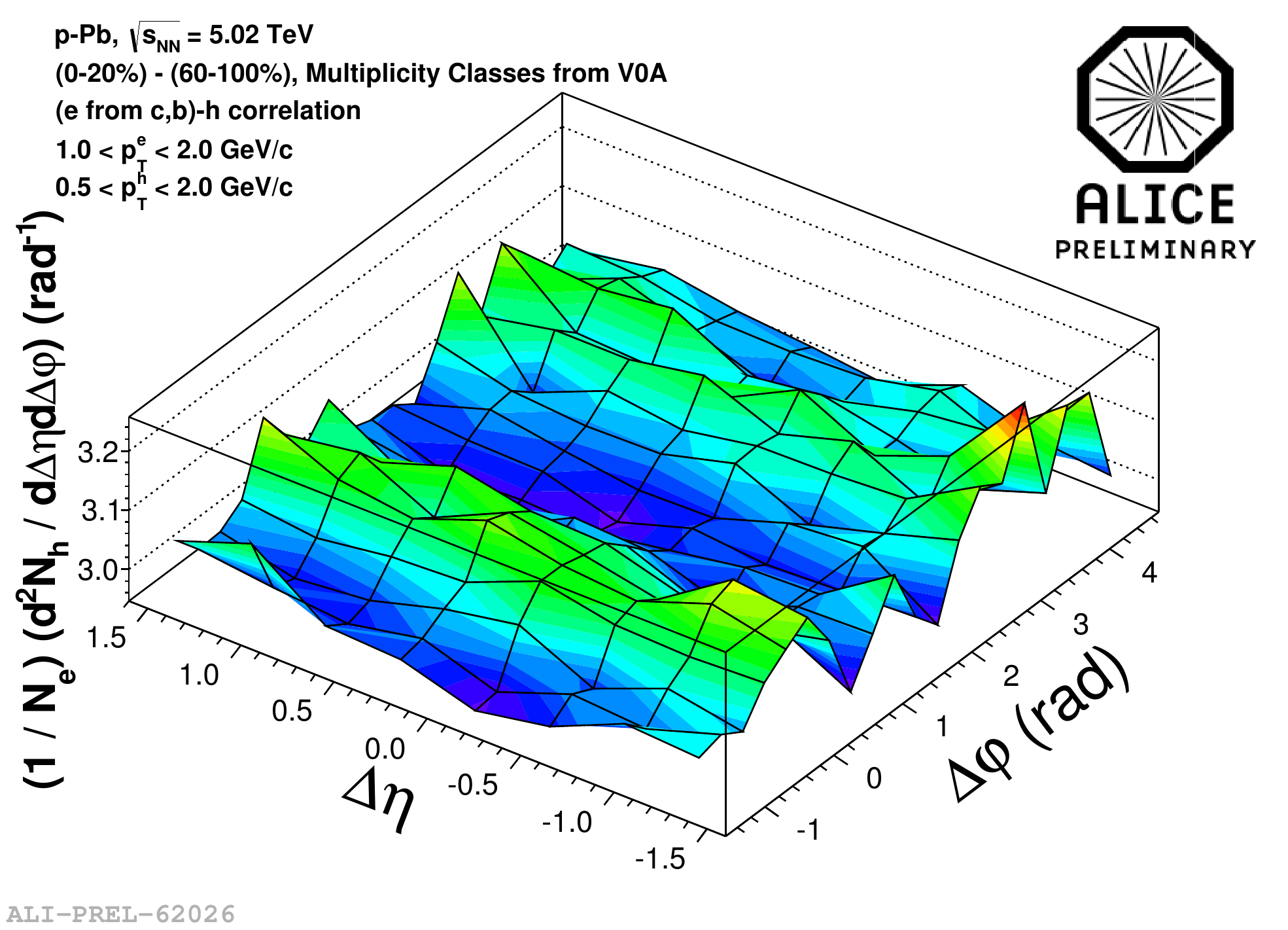}
	} 

\caption{Electron-hadron correlations  in p--Pb collisions at $\sqrt{s_{\mathrm{NN}}}=5.02$ TeV.
	Left: e-h $\Delta \varphi$ correlation in 3 multiplicity classes compared to minimum-bias pp collisions. 
	Right: e-h $\Delta \varphi \Delta \eta$ correlation after subtraction of the low multiplicity class from the high multiplicity class.	
\label{fig:e-h_correlations}
}
\end{center}
\end{figure}

\begin{figure}
\begin{center}
\subfigure{
	\label{fig:e-mu_RAA_central}
	\includegraphics[width=0.410\columnwidth]{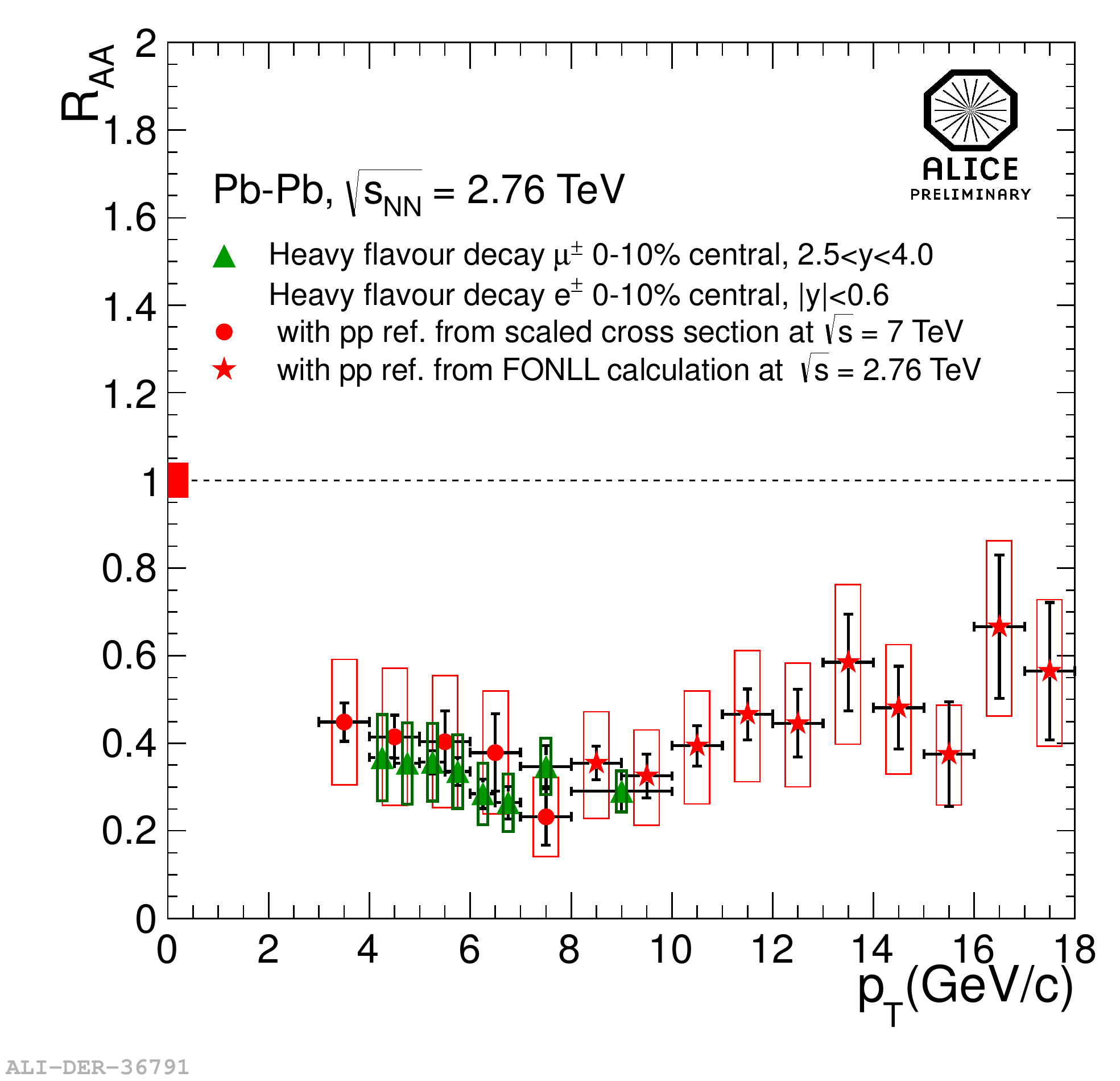}
	} 
\subfigure{
	\label{fig:e-mu_RAA_peripherique}
	\includegraphics[width=0.410\columnwidth]{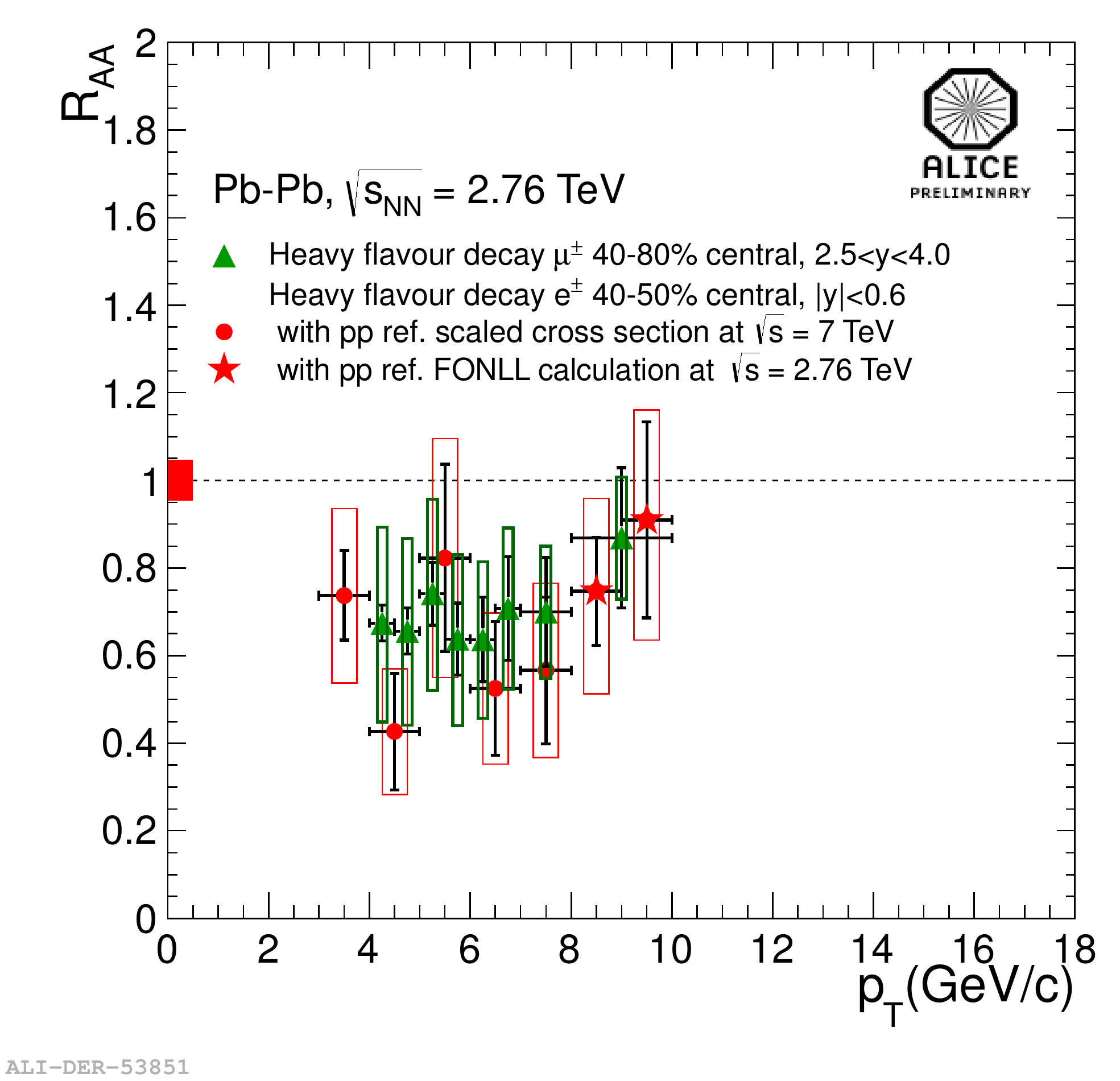}
	} 

\caption{Heavy-flavour decay electron and muon nuclear modification factor $R_{\mathrm{AA}}$ in Pb--Pb collisions at $\sqrt{s_{\mathrm{NN}}}=2.76$ TeV.
	Left: central collisions (0-10\%). 
	Right: semi-peripheral collisions (40-50\% for electrons and 40-80\% for muons \cite{decay_muon_2.76}).	
\label{fig:e-mu_RAA}
}
\end{center}
\end{figure}

\begin{figure}
\begin{center}
\subfigure{
	\label{fig:D_RAA_pt}
	\includegraphics[width=0.455\columnwidth]{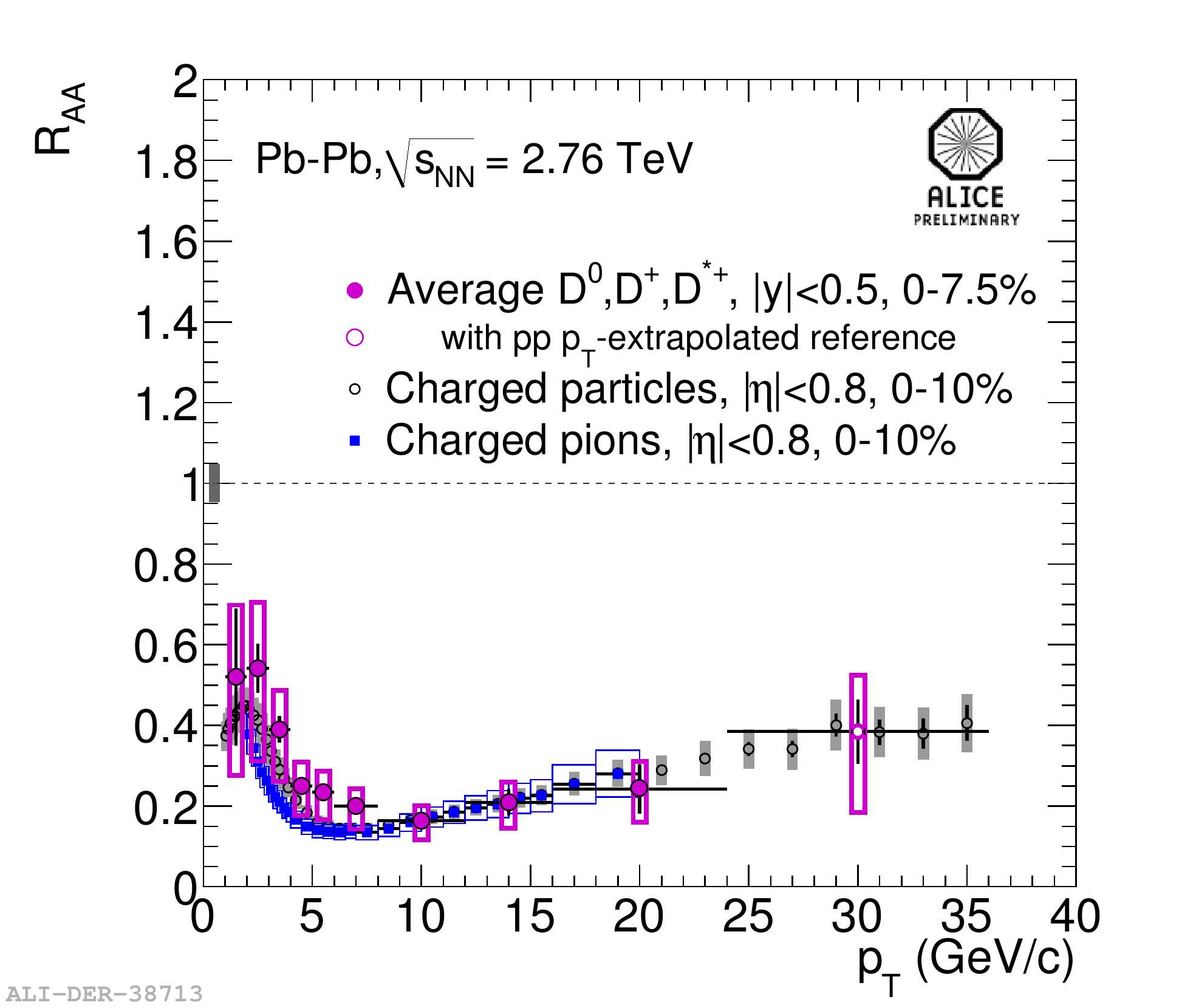}
	} 
\subfigure{
	\label{fig:D_RAA_Npart}
	\includegraphics[width=0.400\columnwidth]{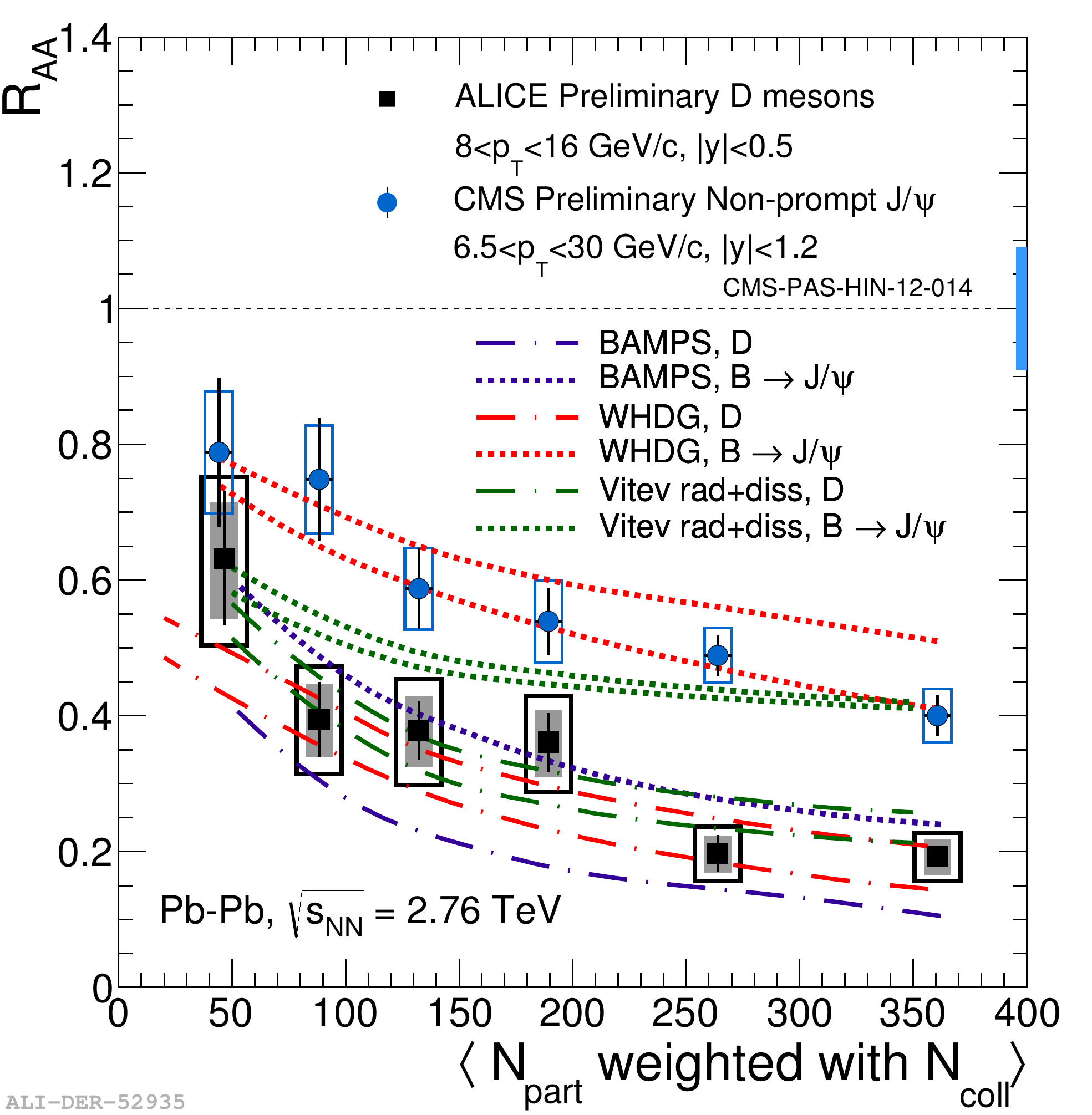}
	} 
\caption{D-meson nuclear modification factor, $R_{\mathrm{AA}}$, in Pb--Pb collisions at $\sqrt{s_{\mathrm{NN}}}=2.76$ TeV.
	Left: $R_{\mathrm{AA}}$ as a function of $p_{\mathrm{T}}$ compared to charged hadrons and pions. 
	Right: $R_{\mathrm{AA}}$ as a function of $N_{\mathrm{part}}$ weighted with $N_{\mathrm{coll}}$ compared to non-prompt J/$\psi$ measured by the CMS collaboration.	
\label{fig:D_RAA}
}
\end{center}
\end{figure}

\begin{figure}
\begin{center}
\subfigure{
	\label{fig:v2_D}
	\includegraphics[width=0.400\columnwidth]{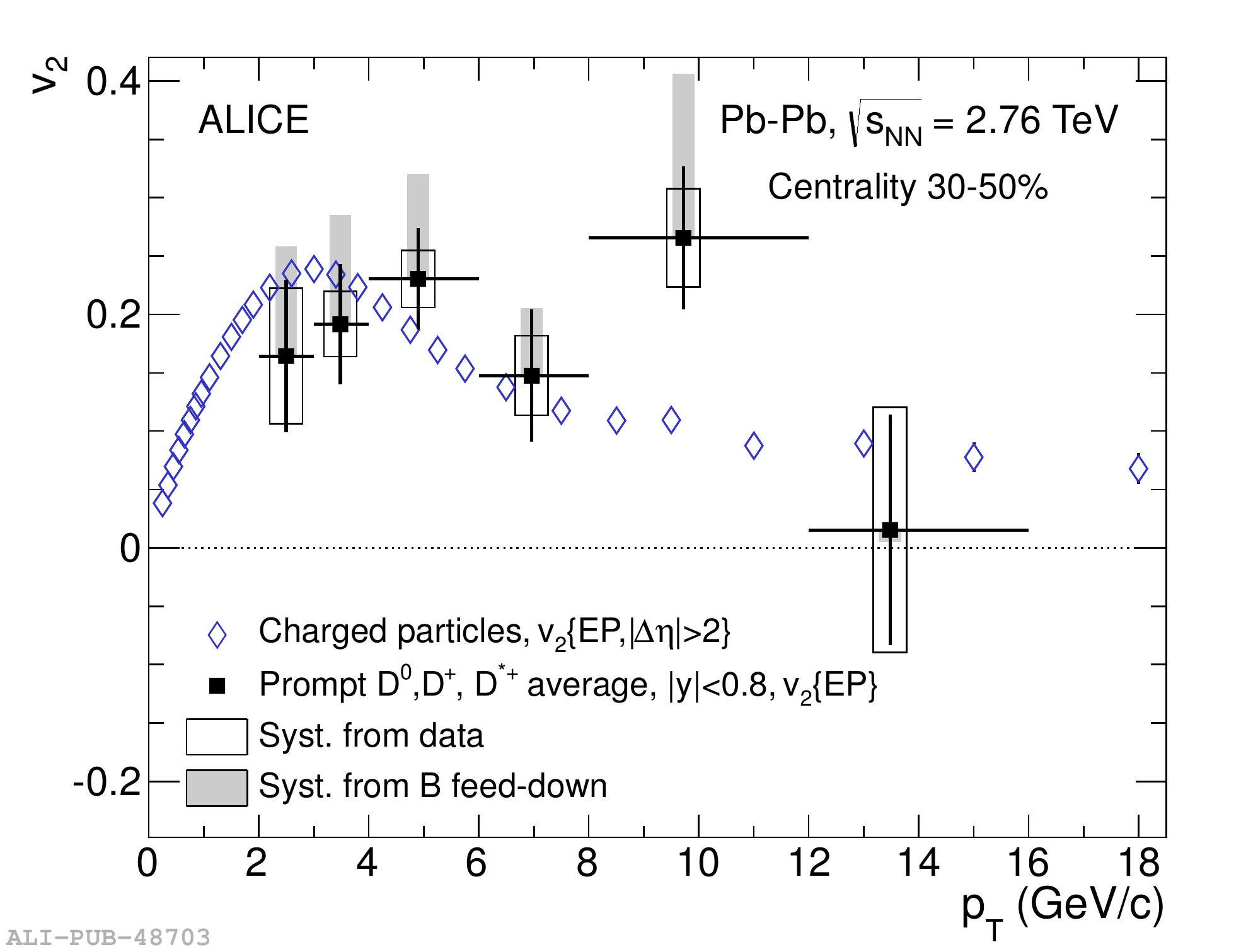}
	} 
\subfigure{
	\label{fig:v2_e_mu}
	\includegraphics[width=0.400\columnwidth]{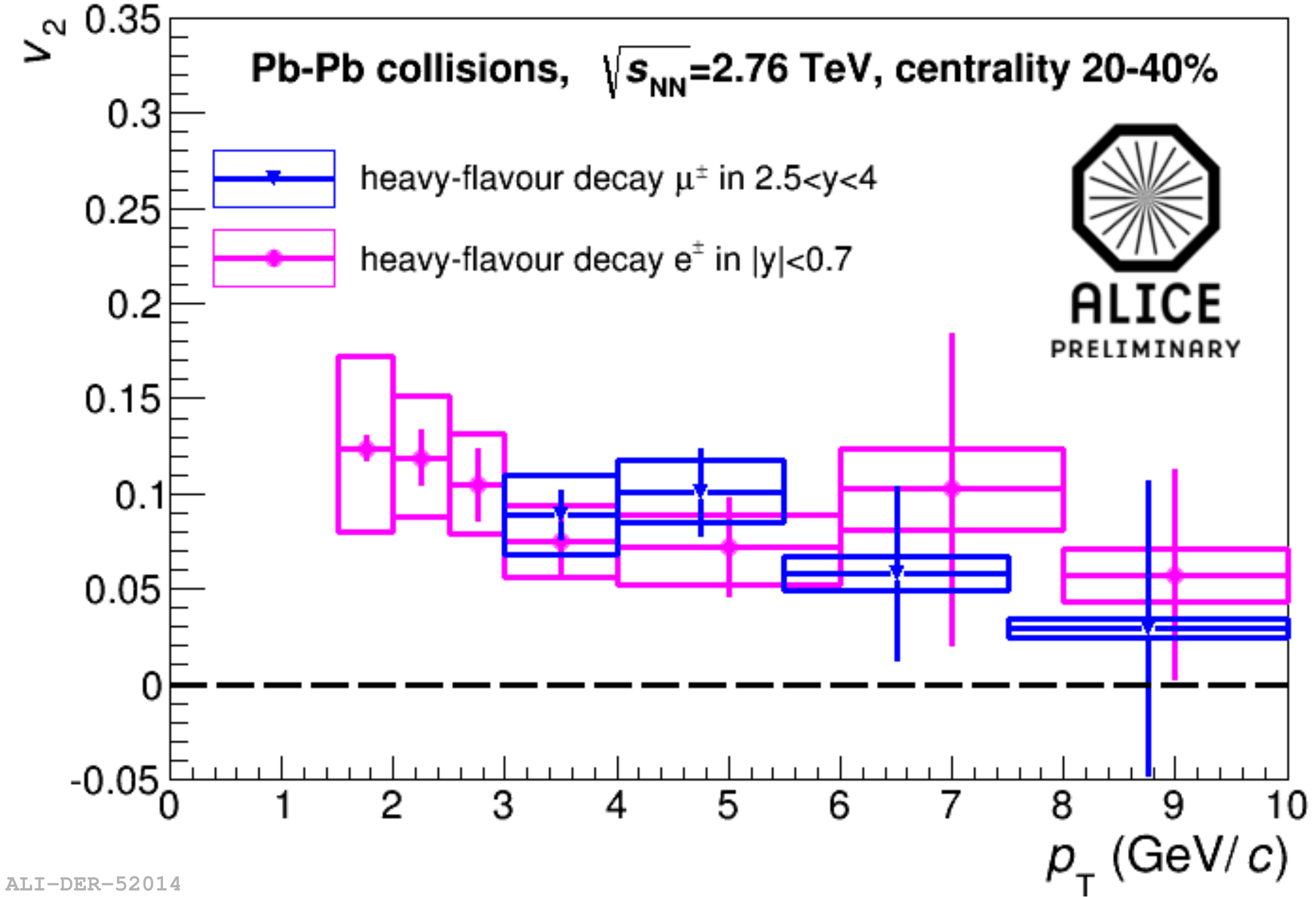}
	} 
\caption{Open heavy-flavour elliptic flow, $v_2$, as a function of $p_{\mathrm{T}}$  in Pb--Pb collisions at $\sqrt{s_{\mathrm{NN}}}=2.76$ TeV.
	Left: D-meson $v_2$ \cite{D_Pb--Pb_v2}. 
	Right: Heavy-flavour decay e and $\mu$ $v_2$.	
\label{fig:v2}
}
\end{center}
\end{figure}

\begin{figure}
\begin{center}
\subfigure{
	\label{fig:D_RAA_theory}
	\includegraphics[width=0.400\columnwidth]{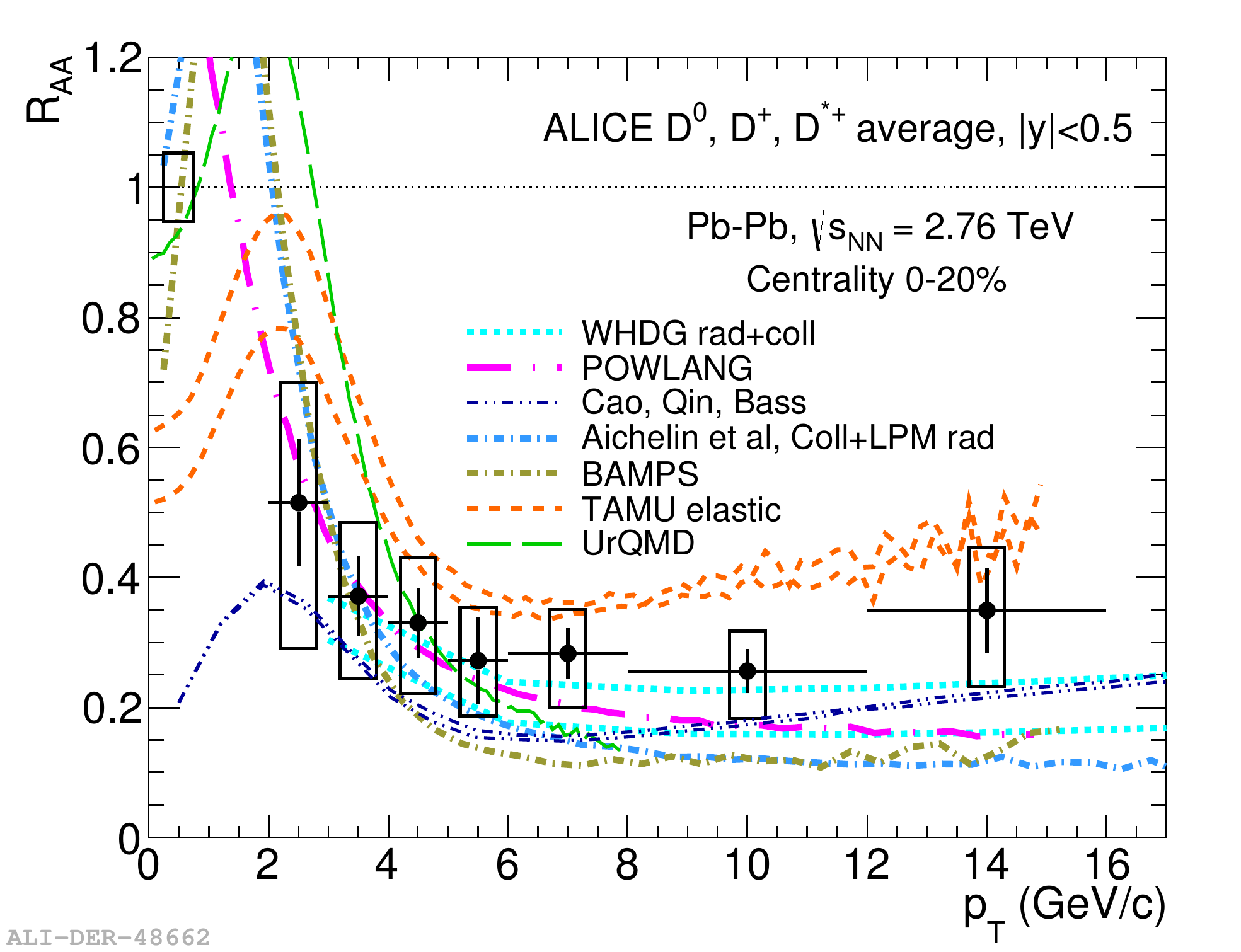}
	} 
\subfigure{
	\label{fig:D_v2_theory}
	\includegraphics[width=0.400\columnwidth]{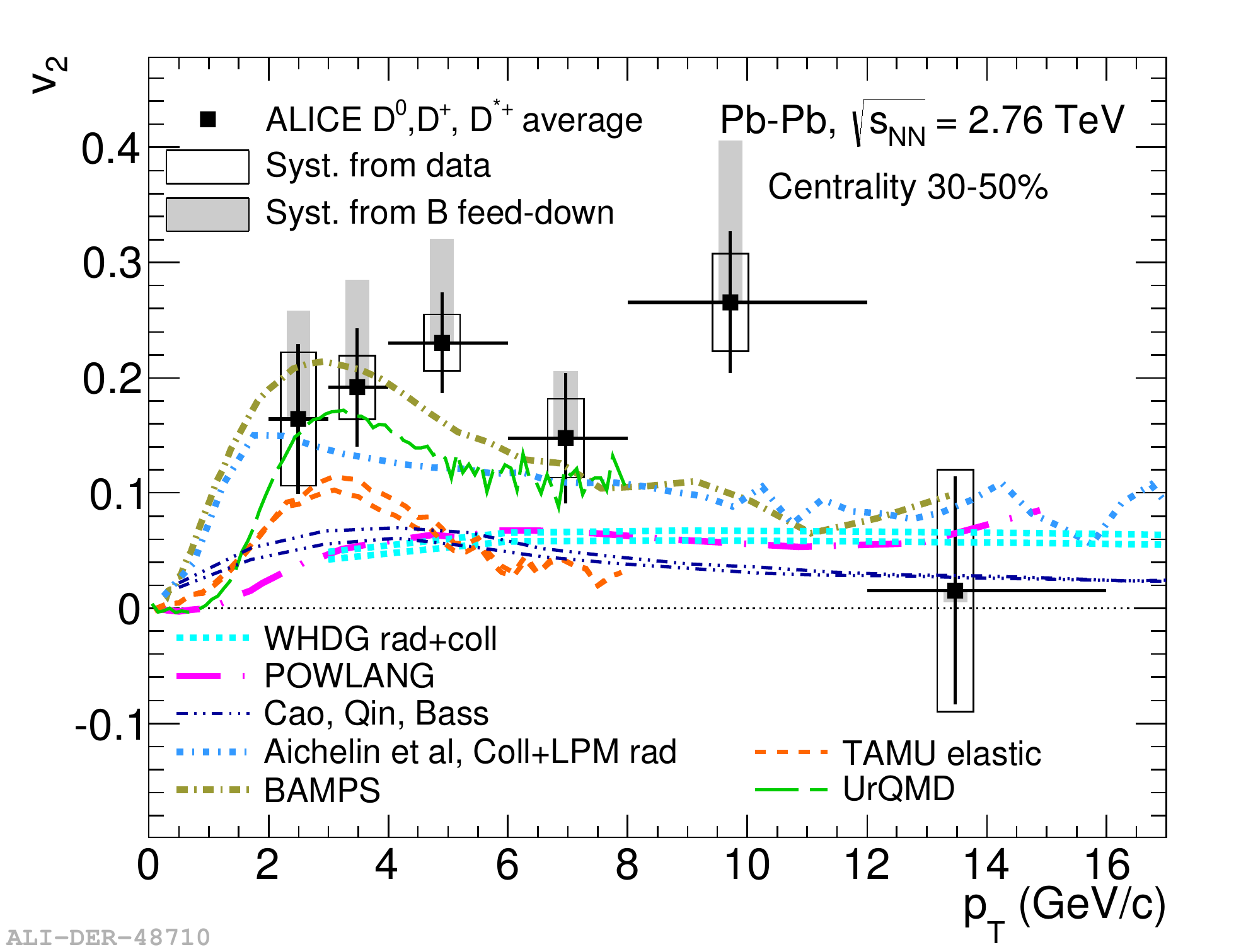}
	} 
\caption{D-meson $R_{\mathrm{AA}}$ and $v_2$ compared to model predictions. Only models with predictions for both $R_{\mathrm{AA}}$ and $v_2$ are shown.
	Left: D-meson $R_{\mathrm{AA}}$  as a function of $p_{\mathrm{T}}$ \cite{D_RAA}. 
	Right: D-meson $v_2$ as a function of $p_{\mathrm{T}}$ \cite{D_Pb--Pb_v2}.	
\label{fig:D_RAA_v2_theory}
}
\end{center}
\end{figure}

\section{Results from p--Pb collisions at $\sqrt{s_{\mathrm{NN}}}=5.02$ TeV}
\label{p--Pb}

 Figure \ref{fig:RpPb} shows the nuclear modification factor $R_{\mathrm{pPb}}$ for D mesons (average of D$^0$, D$^+$ and D$^*_+$) and HF decay electrons.
Both measurements are compatible with unity within statistical and systematic uncertainties. They are consistent with theoretical predictions including a pQCD approach and  EPS09 nuclear modification of the PDFs \cite{MNR, flavour}. The D-meson measurement is also compatible with Color Glass Condensate (CGC) predictions \cite{CGC}. Results are similar for the four D-meson species, including the D$_s^+$ not shown here. The HF-e measurement is in agreement with PHENIX results at lower energy ($\sqrt{s}=0.2$ TeV) \cite{PHENIX_e}.

Figure \ref{fig:e-h_correlations} presents results on angular correlations between HF decay electrons and charged hadrons. The left plot shows the azimuthal angular correlation between trigger particles (HF decay electrons) and associated particles (charged hadrons). The analysis is performed in three multiplicity classes and compared to pp minimum bias results. The highest multiplicity class (0-20\%) presents a stronger correlation than the one observed in the class 60-100\%, which is compatible with pp results. In the right panel of Figure \ref{fig:e-h_correlations} the low multiplicity correlation is subtracted from the high multiplicity one, showing a double-ridge structure as observed in hadron-hadron (h-h) correlations \cite{hh}. This structure can be interpreted in terms of the hydrodynamical evolution of the system, as well as initial conditions originating from CGC.

\section{Results from Pb--Pb collisions at $\sqrt{s_{\mathrm{NN}}}=2.76$ TeV}
\label{Pb--Pb}

Figure \ref{fig:e-mu_RAA} presents the nuclear modification factor $R_{\mathrm{AA}}$  for HF decay electrons and muons \cite{decay_muon_2.76} in central (left) and semi-peripheral collisions (right). A clear suppression is seen for the most central collisions in the explored $p_{\mathrm{T}}$ range. It can be understood as a result of final state effects, $R_{\mathrm{pPb}}$ being compatible with unity (see Figure \ref{fig:RpPb}). A similar suppression is observed at central and forward rapidities with electrons and muons. The pp reference, needed for the $R_{\mathrm{AA}}$ calculation, is obtained from pp data at $\sqrt{s}=2.76$ TeV for HF decay muons, while for D mesons and HF decay electrons a pQCD-scaling of the cross sections measured at$\sqrt{s}=7$ TeV is used \cite{scaling}.
 
The D-meson nuclear modification factor as a function of $p_{\mathrm{T}}$  is presented in Figure \ref{fig:D_RAA}. A suppression up to a factor five is seen at $p_{\mathrm{T}} \sim  10$ GeV/$c$ for the 0-7.5\% most central collisions; again, this can be interpreted as a final satet effect since $R_{\mathrm{pPb}}^{\mathrm{D}}$ is consistent with unity (cf. Figure \ref{fig:RpPb}). One should note that the D-meson nuclear modification factor is similar to that of charged pions and charged particles within uncertainties; however, that the $R_{\mathrm{AA}}$ of D mesons and pions is also sensitive to the shape of the parton energy distribution and their fragmentation functions. Hence, the expected hierarchy in the energy loss $\Delta E$ of c quarks, light quarks and gluons is not trivially linked to the nuclear modification factors of D mesons and pions \cite{D_fragmentation}. In Figure \ref{fig:D_RAA}, $R_{\mathrm{AA}}^{\mathrm{D}}$ as a function of $N_{\mathrm{part}}$ is shown (right). This measurement is compared with results from the CMS collaboration on non-prompt J/$\psi$ and theoretical predictions from \cite{transport, tomography, light_cone}. For D mesons a smaller suppression in peripheral than in central collisions is observed. A larger suppression is seen for D mesons than for non-prompt J/$\psi$, indicating a larger suppression for charm than for beauty. This observation is supported by predictions from energy-loss models, where the difference between the  $R_{\mathrm{AA}}$ of D and B mesons arises from the different masses of c and b quarks ($\Delta E_{\mathrm{c}} > \Delta E_{\mathrm{b}}$).    

Figure \ref{fig:v2} presents the elliptic-flow measurement ($v_2=<cos[2(\varphi-\psi_{\mathrm{RP}})]>$) of D mesons and HF decay electrons and muons. The left panel indicates that the D-meson $v_2$ is similar to that of charged particles within uncertainties. Charm quarks, due to interactions with the medium constituents, acquire properties of the expanding bulk.  Figure \ref{fig:v2} (right) shows that the $v_2$ of electrons and muons from HF decay is positive and compatible within uncertainties in two different rapidity regions. All HF measurements show a positive $v_2$ ($>3\sigma$ effect), indicating that the initial bulk anisotropy is transferred to charm quarks.

$R_{\mathrm{AA}}$ and $v_2$ are two complementary measurements to gain insight into the heavy-quark transport coefficient of the medium. Figure \ref{fig:D_RAA_v2_theory} presents the D-meson $R_{\mathrm{AA}}$ and $v_2$ compared to predictions from seven models \cite{D_RAA,D_Pb--Pb_v2}. It is challenging for models to reproduce simultaneously the two observables at present.   

\section{Conclusion}

With data collected during Run 1, the ALICE collaboration has provided essential measurements for the understanding of heavy-quark production in hadronic collisions and the interaction of heavy quarks with the quark-gluon plasma. D, HF-e and HF-$\mu$ cross sections have been measured in pp collisions at $\sqrt{s}=2.76$ and $\sqrt{s}=7$ TeV. Perturbative QCD calculations describe the measurements within uncertainties. To better understand production processes in the hadronic environment, more exclusive studies are ongoing. Angular correlations between D mesons and charged hadrons integrated over multiplicity are well described by PYTHIA 6.4. D and non-prompt J/$\psi$ yields present an approximately linear increase with the multiplicity of particles produced in the collision. In p--Pb collisions,  D-meson and HF decay electron nuclear modification factors are compatible with unity within uncertainties as predicted by computations including cold nuclear matter effects. In p--Pb collisions, angular correlations between HF decay electrons and charged hadrons exhibit a similar double-ridge structure as that observed in hadron-hadron correlations. The interpretation of this structure is still strongly debated. In Pb--Pb central collisions, a strong suppression at high $p_{\mathrm{T}}$ is seen for D-mesons, HE-e and HF-$\mu$. This is interpreted as evidence for final state effects due to parton energy loss in the hot, dense and colored medium.  The D-meson nuclear modification factor is similar to that of charged pions in central collisions. An ordering is observed for nuclear modification factor of D and non-prompt J/$\psi$ in central collisions compatible with the expectation of a larger in-medium energy loss of charm compared to beauty quarks. A positive $v_2$ is measured in all HF channels at low $p_{\mathrm{T}}$ suggesting that charm takes part in the collective expansion of the medium. High precision heavy-flavour studies are foreseen in next LHC runs.

\end{document}